\documentclass[conference]{IEEEtran}
\IEEEoverridecommandlockouts
\usepackage{cite}
\usepackage{amsmath,amssymb,amsfonts, bm}
\usepackage{algorithmic}
\usepackage{graphicx}
\usepackage{textcomp}
\usepackage[dvipsnames,table,xcdraw]{xcolor}
\usepackage{glossaries} 
\usepackage[numbers, sort]{natbib}
\usepackage[font=small]{caption}

\makeglossaries%
\newacronym{iaq}{IAQ}{indoor air quality}
\newacronym{ldr}{LDR}{light-dependent resistor}
\newacronym{iwast}{IWAST}{IoT With a Soft Touch}
\newacronym{lorawan}{LoRaWAN}{long-range wide-area network}
\newacronym{bsec}{BSEC}{Bosch Software Environmental Cluster}
\newacronym{vocs}{VOCs}{volatile organic compounds}
\newacronym{ulpp}{ULPP}{Ultra Low Power Plus}
\newacronym{ulp}{ULP}{Ultra Low Power}

\newacronym{wos}{WoS}{wake-on-sound}
\newacronym{adc}{ADC}{analog-to-digital converter}
\newacronym{wdt}{WDT}{watchdog timer}
\newacronym{mems}{MEMS}{microelectromechanical systems}
\newacronym{lipo}{LiPo}{lithium polymer}
\newacronym{pcl}{PCL}{problem-centered learning}
\newacronym{iot}{IoT}{internet-of-things}
\newacronym{aqi}{AQI}{air quality indicator}

\usepackage{tikz}
\usetikzlibrary{external}
\usepackage{pgfplotstable}
\usepgfplotslibrary{groupplots,dateplot}
\usetikzlibrary{patterns,shapes.arrows}
\pgfplotsset{compat=newest}
\usetikzlibrary{spy}
\usepgfplotslibrary{polar}
\usetikzlibrary{positioning}
\usetikzlibrary{plotmarks}
\usetikzlibrary{arrows}
\usetikzlibrary{arrows.meta}
\usepgfplotslibrary{patchplots}
\usetikzlibrary{decorations, decorations.text,backgrounds} 
\usepgfplotslibrary{colorbrewer}
\usetikzlibrary{pgfplots.statistics}
\usepackage[binary-units=true]{siunitx}

\usepackage{threeparttable}
\usepackage{booktabs}
\usepackage{url}
\usepackage[binary-units=true]{siunitx}
\DeclareSIUnit{\Wh}{Wh}
\DeclareSIUnit{\belmilliwatt}{Bm}
\DeclareSIUnit{\dBm}{\deci\belmilliwatt}
\DeclareSIUnit{\belmillii}{Bi}
\DeclareSIUnit{\dBi}{\deci\belmillii}

\definecolor{color0}{rgb}{0.12156862745098,0.466666666666667,0.705882352941177}
\definecolor{color1}{rgb}{1,0.498039215686275,0.0549019607843137}
\definecolor{color2}{rgb}{0.172549019607843,0.627450980392157,0.172549019607843}

\usepackage{subcaption}

\usepackage{comment}

\usepackage{hyperref}

\usepackage{pgfplots}
\usetikzlibrary{pgfplots.groupplots}
\usepackage{lipsum,adjustbox}

\newcommand{\ie}{{i.e.,\ }}
\newcommand{\eg}{{e.g.,\ }}

\pgfplotsset{every axis/.append style={
                    label style={font=\footnotesize},
                    tick label style={font=\footnotesize}  
                    }}

\def\BibTeX{{\rm B\kern-.05em{\sc i\kern-.025em b}\kern-.08em
    T\kern-.1667em\lower.7ex\hbox{E}\kern-.125emX}}

\begin{document}

\title{IoT with a Soft Touch: A Modular Remote Sensing Platform for STE(A)M Applications}

\author{\IEEEauthorblockN{Jona Cappelle,  Geoffrey Ottoy, Sarah Goossens, Hanne Deprez,\\ Jarne Van Mulders, Guus Leenders, Gilles Callebaut}

\IEEEauthorblockA{KU Leuven, ESAT-WaveCore, Ghent Technology Campus, 9000 Ghent, Belgium}
}

\maketitle


\begin{abstract}
Besides wide attraction in the industry, IoT is being used to advance STEM and STEAM education across a range of education levels. This work presents a remote sensing platform, named \textit{IoT with a Soft Touch}, developed to achieve two goals. First, it aims to lower the technicality, stimulating the students to do STE(A)M. Second, the technology is to be used in ``softer'' applications (e.g., environmental and health care), thereby aiming to attract a more diverse set of student profiles. 
Students can easily build a wireless sensing device, with a specific application in mind. The modular design of the platform and an intuitive graphical configurator tool allows them to tailor the device's functionality to their needs. The sensor's data is transmitted wirelessly with LoRaWAN. The data can be viewed and analyzed on a dashboard, or the raw data can be extracted for further processing, e.g., as part of the school's STE(A)M curriculum. This work elaborates on the low-power and modular design challenges, and how the platform is used in education.
\end{abstract}


\section{Introduction}

The \emph{ROSE} study~\cite{sjoberg2010rose} showed that pupils hold positive attitudes towards science and technology, but skepticism towards STEM is growing amongst youngsters, especially in the richest countries, such as Northern Europe and Japan. Furthermore, girls are more negative or ambivalent towards the role of science and technology in our society than boys. Especially in developed countries, youngsters are not enthusiastic about school science, do not believe that school science will advance their careers, or even think that school science has opened their eyes to new and exciting jobs. Large gender differences exist in the STEM topics that interest boys and girls: where boys are more attracted towards technical, mechanical, electrical topics and spectacular experiments, girls are mainly interested in health and medicine, the human body, ethics, and societally relevant topics. As studied in~\cite{botella2019gender}, to attract more girls/women to STEM, we need to increase the visibility of areas that are attractive to the female gender and build a welcoming environment to achieve a long-term change. This was one of the key principles behind the presented platform.

\begin{figure}[t]
    \centering
    \includegraphics[width=0.9\linewidth]{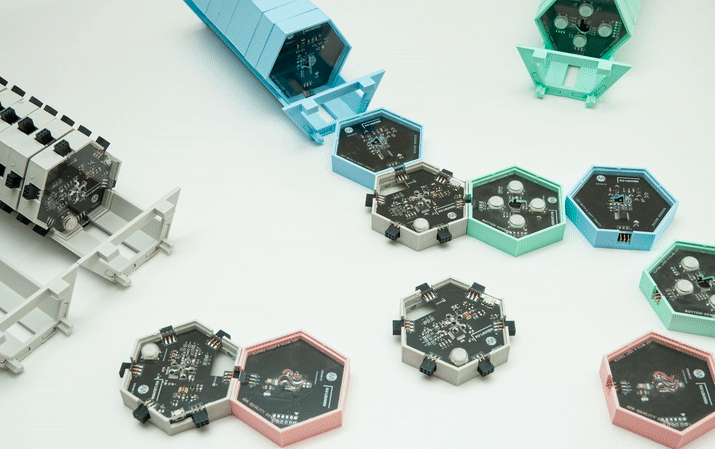}
    \caption{Several IWAST sensor modules (colored) and motherboards (gray).}%
    \label{fig:system}
\end{figure}

\Gls{iot} provides new possibilities and opportunities to acquire a certain skill-set and knowledge, and therefore has a crucial role to play in education, as elaborated in ~\cite{TIoT, tae2017development}. We designed an open-ended \gls{iot} STEM project according to integrated STEM instructional practices~\cite{thibaut2018integrated}, especially incorporating \gls{pcl}, integration between STEM disciplines (INT), modeling (MOD), and inquiry- and design-based learning (IBL and DBL).

In an effort to show the societal relevance of science and technology and to enhance their research skills, students are challenged to investigate an ecologically or societal relevant problem in their school environment with our plug-and-play IoT set (PCL). This paper will focus on the design and technical challenges when making the \gls{iot} set and platform (Fig.~\ref{fig:system}). 
Sensor metrics (\ie temperature, air pressure, air humidity, air quality, sound level, light intensity, and button presses) were selected such that students can study sound pollution and environmental conditions in their everyday surroundings.
The chosen IoT wireless communication technology, LoRaWAN, poses important preconditions, such as maximum signal range, transmission frequency, etc. for the students' chosen problem statement.
Teachers can elaborate on mathematical topics ((statistical) data processing), technical topics (wireless communication, sensor technology), science topics (problem statement, sound, resistive sensor technology) (INT). 
Based on their self-formulated research questions (IBL), students are encouraged to select the appropriate sensor metrics and configuration (DBL) and report appropriate conclusions after data collection and processing (IBL).
The design team focused on designing and assembling IoT sensor modules and configuring the wireless communication link, as high school students generally do not have the necessary technical baggage.


The remainder of the paper is structured as follows. First, an overview of the full IWAST system is given, and the individual components are elaborated. In Section~\ref{sec:low-power-design}, we describe the approaches taken to increase the autonomy of the remote sensing modular platform. Two applications studied by the students are described in Section~\ref{sec:applications}. The conclusions and the future work are presented in Section~\ref{sec:conclusions}.

\section{IWAST Platform}\label{sec:iwast-platform}
\begin{figure}[t]
    \centering
    \includegraphics[width=0.55\linewidth]{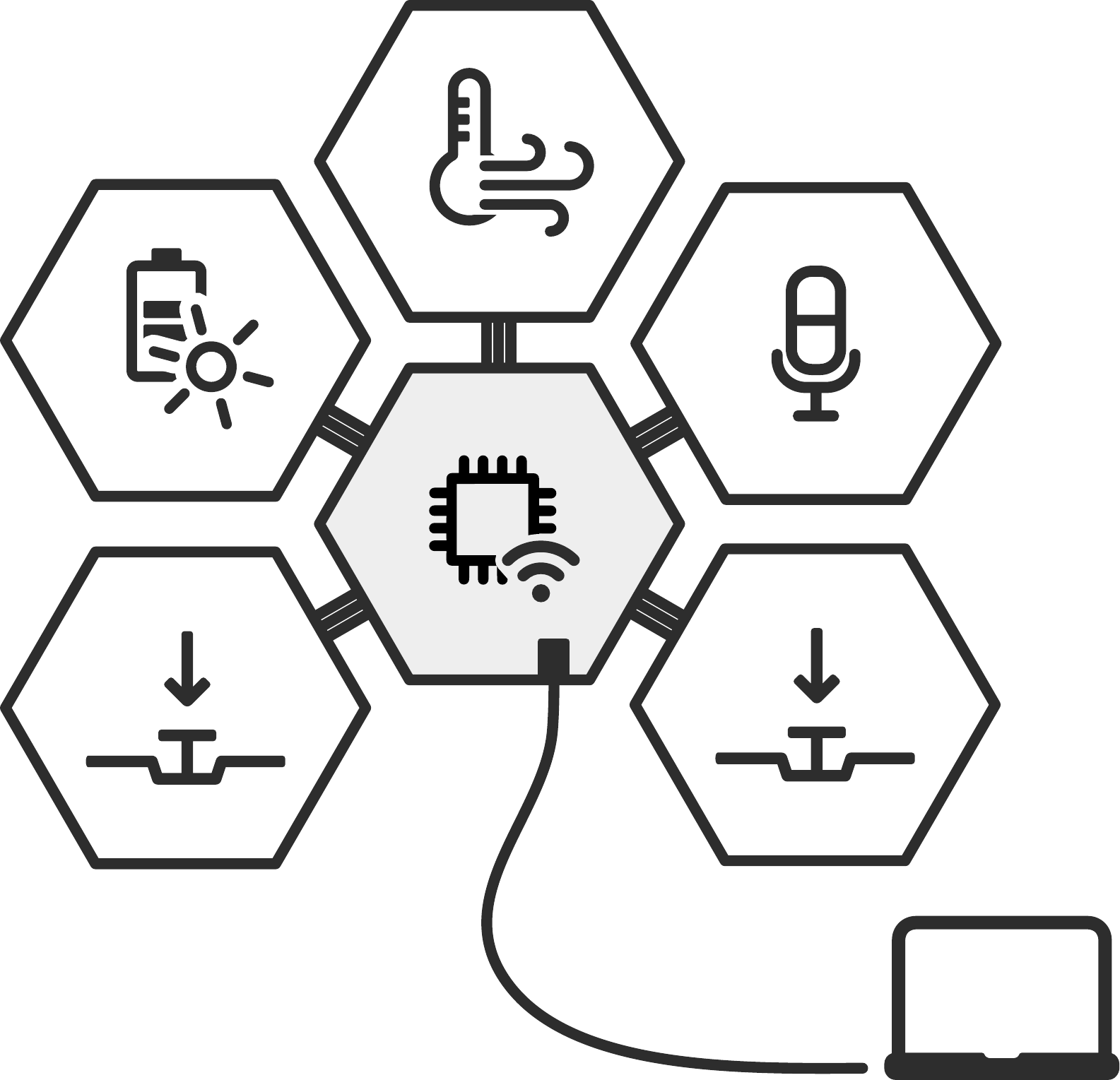}
    \caption{Schematic overview of the \gls{iwast} hardware platform. The main controller board (motherboard) controls all connected sensor boards and communicates with the cloud. Each set is programmable via intuitive software on a computer. 
    }%
    \label{fig:systemicons}
\end{figure}

The presented \gls{iwast} platform consists of both hardware modules and supporting software packages. This is illustrated in Fig.~\ref{fig:systemicons}.
The hardware platform consist of several hexagon-shaped boards. The central board, is the dedicated main controller, i.e., \emph{the motherboard}. The other hexagon-shaped boards (\emph{the sensor boards}) can be connected to all faces of the motherboard. 
The motherboard reads all connected sensors and wirelessly connects to the cloud. The internal processing of an \gls{iwast} set can be adjusted to a specific use case via the dedicated configurator on a computer via USB. There, depending on the connected sensor boards, different parameters can be configured. For example, only when a sensor value drops below or exceeds a certain level (\ie threshold), the sensor value will be transmitted. In contrast, the sensor module can be requested to read out its sensor(s) at specific intervals. The choice between these distinct measuring operations, \ie polling- and threshold-based measuring, has a large impact on the energy consumption, which is discussed in Section~\ref{sec:low-power-design}.

After the configuration stage, the sensor measurements are sent wirelessly over LoRaWAN via \textit{The Things Network}. On our cloud platform, all measurements are stored in a local database and can be accessed through our \textit{IWAST dashboard}. There, students have a clear overview of all sensor data. In addition, the raw sensor measurements can be downloaded for further offline processing. 
In what follows, the different components in the IWAST system are elaborated in more detail.

\subsection{Motherboard}
The \gls{iwast} motherboard, the central controller of an \gls{iwast} set, automatically discovers and connects to all connected sensor boards. The motherboard features a built-in USB connection, which makes for easy communication with our accompanying configuration software. There, all use case parameters (\eg polling interval, thresholds) can be configured.  The motherboard is equipped with a LoRaWAN radio to connect a \gls{iwast} set to the cloud.  
The motherboard can be powered from either the computer's USB connection or by the power sensor board featuring an internal battery.

\subsection{Sensor Boards}
A distinction is made between the sensor, the sensor board, and the sensor metric. A sensor board comprises one or more sensors, which in turn collects one or more sensor metrics. To give an example, the environmental sensor board features one sensor, measuring several sensor metrics \eg atmospheric pressure, temperature, and relative humidity.


Currently, four different sensor boards have been developed: a) an environmental sensor, b) a microphone sensor, c) a button sensor, and d) a power module.   
The \emph{environmental sensor board} accurately measures parameters regarding the environment: i) temperature, ii) air pressure, iii) \gls{iaq} index and iv) relative (air) humidity. The \gls{iaq} index is determined based on both the current measured and previous sensor values. It indicates or quantifies the quality of the air available in the surrounding. The \emph{microphone sensor board} measures the audio level of the surroundings using a microphone by collecting a \SI{20}{ms} audio recording.
The \emph{button sensor board} hosts four buttons with four accompanying LEDs. When a button is pressed, all buttons will briefly light up.  The motherboard and sensor boards can be powered through the \emph{power module board}. It integrates a battery with an ambient light energy harvester.
Next to this, it also hosts a light sensor, measuring the light level of the surroundings. If for any reason the battery gets drained, there is still a possibility to charge the battery using the micro-USB port.

\subsection{Dashboard}
Students can log in on the dashboard\footnote{\url{dramco.be/projects/iwast/platform}} with their unique session name. The displayed sensor data can be selected from different periods, i.e., day, week, month, year, all. The dashboard includes a graph for each type of sensor metric. Each motherboard has a distinct coloring, to conveniently see the sensor values originating from the same motherboard.

\section{Modular and Low-Power Design}\label{sec:low-power-design}
The boards are designed taking good practice~\cite{s21030913} in mind to keep the energy consumption low. All \gls{iwast} documentation and source code files (hardware and firmware) are open-source and available online\footnote{\url{github.com/dramco-iwast}}. The power consumption of the separate sensor boards is summarized in Table~\ref{tab:power}.



\subsection{Motherboard}
To lower the technical skills required to build a wireless sensing device, we have designed the motherboard in a modular and versatile way.
Sensor boards can be plugged into each of the six sides of the \gls{iwast} motherboard. To standardize the communication between motherboard and sensor board, an I2C bus interface is used, implemented with one custom command set across all sensor boards.\footnote{The sensor command set is described in \url{dramco-iwast.github.io/docs}.} By decoupling the motherboard from the sensor-specific communication, we provide a future-oriented interface allowing a possible extension. To that end, each sensor board features a 
microcontroller, which handles on-board sensor configuration, readings, and communication with the motherboard. As such, it acts as a bridge between the I2C communication with the motherboard and any sensor-specific interface, \eg analog, SPI, etc.


To define the desired behavior for a specific set, \ie a motherboard with its connected sensors, the motherboard can be connected via USB to a PC and configured using a custom configurator tool.
The motherboard starts with notifying the configurator of its connected sensors. The behavior for each sensor can be configured separately. For example, using the configurator, we can decide that a first sensor should be read every 5 minutes, while a second sensor will only notify the motherboard if its readings are above a certain threshold. All these settings, along with communication-specific settings, e.g., security keys, are stored in the motherboard's non-volatile memory.

When a motherboard is powered (or on reset), it will wait for 30~seconds for a USB connection. After that, it will configure its connected sensors according to the configuration stored in its non-volatile memory. After that, it enters a low-power state.

During low-power operation (sleep), the motherboard will periodically wake up and poll sensors for readings based on the configured poll interval and go back to sleep. Whenever a sensor board has data ready, either because a configured threshold is exceeded or because it has been polled earlier, it notifies the motherboard using its dedicated interrupt line. This interrupt will wake up the motherboard, after which it reads the sensor's readings and relays them to the radio for transmission.
Then it's nap time again.


\begin{table}[tbp]
    \centering
    \caption{Power consumption of the sensor modules and motherboard in \si{\micro\ampere} at \SI{3.3}{\volt}.}
    \label{tab:power}
    \begin{tabular}{@{}lrrrr@{}}
    \toprule
    Sensor Module & \multicolumn{2}{c}{Active} & \multicolumn{1}{c}{Inactive} & \multicolumn{1}{c}{Sleep}\\
    \cmidrule(lr){2-3}
    & Current & Time (\si{\milli\second})& & \\
    \midrule
        Motherboard \footnotemark    & 25400 & 7000      & 55            & 55              \\
        Environmental   & 8430  & 3520      & 1060          & -              \\
        Microphone      & 4000  & 500       & 25            & 1             \\
        Button          & 7300  & 2000      & 0.330         & 0.330         \\ 
        Power/Light     & 4000  & 28        & 3.2           & 3.2           \\ \bottomrule
    \end{tabular}
\end{table}

\footnotetext{\acrshort{lorawan} transmission of (typical) accumulated 36 byte data packet using Spreading Factor 11}

\subsection{Environmental Sensor}
The air quality sensor board is equipped with a \textit{Bosh BME680} sensor, accompanied by an \textit{ST STM32L072KZ} microcontroller. 
This controller is running the \gls{bsec} supported firmware\footnote{\url{bosch-sensortec.com/software-tools/software/bsec/}}, running a proprietary algorithm to obtain a calibrated \gls{iaq} value. In essence, this output is an index that can have values between 0 and 500 with a resolution of 1 to indicate or quantify the quality of the air available in the surrounding.

To minimize power consumption, it always runs in a low power state, in which a measurement will only be taken every 5 minutes (\gls{ulp} cycle).
When needed, the motherboard can request extra measurements in between this 5-minute \gls{ulp} cycle. 
Further intelligence is implemented in the firmware to minimize the number of measurements, thus also minimizing power consumption. For example, when the motherboard manually requests a measurement close to an \gls{ulp} cycle, this value is used instead of starting an extra measurement.
When no measurement is going on, the sensor is put to sleep, consuming \SI{1}{\micro A}. The on-board controller remains active, however: consuming \SI{1.06}{mA}, which could be further reduced by letting the on-board controller sleep between the \gls{bsec} cycles.
When sensing, the majority of the consumed power can be attributed to the inherent workings of the gas sensor. This involves two steps:
\begin{enumerate}
    \item Heating the gas sensor hot plate to a target temperature (typically between \SI{200}{^\circ C} and \SI{400}{^\circ C}) and maintaining this temperature for a certain duration of time. This consumes \SI{14}{mA} for a duration of \SI{1.71}{s}. 
    \item Measuring the resistance of the gas-sensitive layer, yielding a consumption of \SI{1.57}{mA} for a duration of \SI{1.85}{s}.
\end{enumerate}
	




\subsection{Microphone Sensor}

The used \textit{Vesper VM1010} \gls{mems} microphone features extremely low active mode current, and includes an efficient sleep mode function. 
The small-signal, originating from the microphone, needs to be filtered and amplified. First, a high pass filter removes DC offsets and amplifies the signal by \SI{30}{dB}. Secondly, any unwanted ultrasonic sounds and harmonics are suppressed by a Sallen-Key low pass filter, designed with a \SI{-3}{dB}-frequency of \SI{8}{kHz}. An amplifier, designed with a gain of \SI{2.73}{dB}, amplifies the resulting signal to an appropriate level for the \gls{adc}.

This board supports \gls{wos} detection, enabling long periods of sleep between sound events, thus optimizing energy consumption drastically. In hardware, three~\gls{wos} levels are available: \si{65}, \si{77}, and \SI{89}{dBSPL}. The level is controlled by a dual-channel analog switch (\textit{Texas Instruments TS5A2066DCUR}) through a resistor network. The users can choose any \gls{wos} level (threshold) between \num{65} and \SI{100}{dBSPL}. The software level is mapped to one of the three hardware \gls{wos}~levels by selecting the maximum hardware \gls{wos}~levels, being lower than the chosen software threshold.  

The sound module supports both interrupt-based (\gls{wos}) and polling-based operation. On a microphone wake-up, either from a \gls{wos} event or a polled measurement, a short audio clip is sampled (400 samples at \SI{20}{kHz} i.e., a duration of \SI{20}{ms}) from which a sound level (in dBSPL) is calculated. Whenever a threshold-value is exceeded, i.e., the calculated sound level is greater than the threshold setting, an interrupt will inform the motherboard of a new measurement. When in polling-mode, the power supply is only enabled for \SI{20}{ms} to further reduce the energy consumption. Note that this will not work in the interrupt mode, as the sensor needs to be powered to provide \acrlong{wos}. To avoid continuous triggering of the thresholds, the microphone is disabled for \SI{1}{min} after a \gls{wos} event. 

As illustrated in~\cite{s21030913}, depending on the periodicity of the events, it is sometimes more energy-efficient to use the polling-mode where we can power-off the sensor, opposed to using the \gls{wos} where the sensor needs to be continuously powered.



\subsection{Button Sensor}
The button-sensor works completely interrupt-based and is always in sleep. Sleep current is below \SI{1}{\micro\ampere}. On a button press, the device becomes active for \SI{1.7}{\second} with an average current consumption of \SI{7.3}{mA}.

\subsection{Power Module}
The \gls{iwast} system is energy provisioned by the power module. This sensor module includes a battery for energy storage, a small solar panel for energy harvesting, and an integrated light sensor.

A single cell \gls{lipo} battery with a capacity of \SI{500}{\milli\ampere h} battery is included in the power module. This battery can be charged by USB and trickle charged through solar energy harvesting. To make the power module energy efficient, the onboard \gls{adc} and voltage divider for measuring the battery voltage is only enabled sporadically. 

A \textit{Panasonic AM-5412} amorphous solar panel enables solar energy harvesting. Combined with a nano power boost charger and buck converter chip (\textit{Texas Instruments BQ25570}), we are able to efficiently extract microwatts to milliwatts of power generated from the photovoltaic cells. 

The power module is equipped with a high accuracy and low power ambient light sensor: the \textit{Vishay VEML7700} (consuming only \SI{6.6}{\micro\watt}). Unfortunately, this sensor does not feature any built-in threshold detection, so a sporadic polling technique is implemented: briefly waking up the system every \SI{16}{\second}, checking the last value of the light sensor and comparing it against the threshold value.

\section{IWAST in the Wild}\label{sec:applications}

\subsection{Ventilation requirements during Covid-19}
The IWAST system was used to evaluate the air quality in classrooms~\cite{TurunenMari2014Ieqi}, especially during the Covid pandemic. Not only was regular ventilation imposed by the government, also there exists a relationship between the air quality and the Covid-19 deaths~\cite{cole2020air}. Therefore, measuring the \gls{iaq} index is imperative for public health. 

\subsection{Noise Levels in the Classroom, Restaurant, and Playground}
Pupils used the IWAST system to measure sound levels in their classroom and playground, with research questions such as `What is the difference in sound level between the front and the back of the classroom?', `How does the sound level measured with interrupt-based IoT sensing compare to the sound level measured with a sound level meter?', and `Which class visiting a classroom is the noisiest (and thus likely the least attentive \cite{ShieldBridgetM2008Teoe})?'.
Furthermore, pupils also plan to investigate the relation between classroom air quality with sound level, assuming a class to be noisier and less attentive when air quality is worse.

\section{Conclusions and Future Work}\label{sec:conclusions}
This paper describes the design of a modular remote sensing platform for STE(A)M applications, called \emph{\acrlong{iwast}}. This \gls{iot} platform is developed to i) stimulate students to do STE(A)M by lowering the technicality and ii) targeting `softer' applications (\eg environmental and health care).
The system comprises a motherboard and some sensor boards. Next to an elaboration of the \gls{iwast} system, we explain how we addressed the challenges related to the design of a modular and low-power platform. The modularity and flexibility of the system are obtained by using hexagon-shaped boards, allowing to connect a sensor to all faces of the motherboard. Moreover, the use of dedicated microcontrollers on the sensor boards decouples the inner-workings of the sensors from the motherboard, making it easily extendable. The platform employs strategies such as sleep-modes and interrupt-based communication to extend the battery lifetime as much as possible.
Based on the evaluation and the needs of the schools, we pursue to further extend the set of different sensor boards. In addition, the concepts to improve the energy efficiency of these systems will be made more clear towards the pupils. For instance, we will include a power report to illustrate the effect on the energy consumption based on the configured parameters.

\section*{Acknowldegments}
The authors would like to thank Bart Thoen and Matthias Alleman for their help in designing the first prototypes.

{\footnotesize%
\bibliographystyle{IEEEtranN}%
\bibliography{./bib}%
}

\clearpage

\end{document}